\documentclass[aps,twocolumn,showpacs]{revtex4-1}

\usepackage{graphicx}

\begin{document}

\title{Decay of energy and suppression of Fermi acceleration in a
dissipative driven stadium-like billiard}

\author{Andr\'e L.\ P.\ Livorati$^1$, Iber\^e L.\ Caldas$^1$ and Edson
D.\ Leonel$^2$}

\affiliation{$^1$Instituto de F\'isica - IFUSP - Universidade de S\~ao
Paulo - USP\\ Rua do Mat\~ao, Tr.R, 187 - Cidade Universit\'aria --
05314-970 -- S\~ao Paulo -- SP -- Brazil\\ $^2$Departamento de
Estat\'istica, Matem\'atica Aplicada e Computa\c c\~ao, UNESP -- Univ
Estadual Paulista\\ Av.24A, 1515 -- Bela Vista -- 13506-900 -- Rio Claro
-- SP -- Brazil}

\date{\today} \widetext


\begin{abstract}
The behavior of the average energy for an ensemble of non-interacting
particles is studied using scaling arguments in a dissipative
time-dependent stadium-like billiard. The dynamics of the system is
described by a four dimensional nonlinear mapping. The dissipation is
introduced via inelastic collisions between the particles and the moving
boundary. For different combinations of initial velocities and damping
coefficients, the long time dynamics of the particles leads them to reach
different states of final energy and to visit different attractors,
which change as the dissipation is varied. The decay of the
average energy of the particles, which is observed for a large range of
restitution coefficients and different initial velocities, is described
using scaling arguments. Since this system exhibits unlimited energy
growth in the absence of dissipation, our results for the dissipative
case give support to the principle that Fermi acceleration seem not to
be a structurally stable phenomenon.
\end{abstract}

\maketitle

\textbf{Some dynamical properties of a dissipative time-dependent
stadium-like billiard are studied. The system is described in terms of a
four-dimensional nonlinear mapping. Dissipation is introduced via
inelastic collisions of the particle with the boundary, thus implying
that the particle has a fractional loss of energy upon collision. The
dissipation causes substantial modifications in the dynamics of the
particle as well as in the phase space of the non-dissipative system. In
particular, inelastic collisions are an efficient
mechanism to suppress Fermi acceleration of the particle. We show that a
slight modification of the intensity of the damping coefficient yields
 a change of the final average velocity of the ensemble of particles.
Such a difference in the final plateaus of average velocity is explained
by a large number of attractors created in the phase space by the
introduction of dissipation in the system. We also described the
behavior of decay of energy via a scaling formalism using as variables:
(i) initial velocity ($V_0$); (ii) the damping coefficient ($\gamma$)
and; (iii) number of collisions with the boundary ($n$). The decay of
energy, leading the dynamics to converges to different plateaus in the
low energy regime, is a confirmation that inelastic collisions do
indeed suppress Fermi acceleration in two-dimensional time-dependent
billiards.
}

\section{Introduction}
\label{sec1}
Billiard problems symbolize the dynamics of a point-like particle
(the billiard ball) which moves inside a compact region $Q$,
that in the context of the mathematics of billiards is known as a
billiard table (or for short, billiard). Inside the billiard, the
particle moves in straight lines until it reaches the boundary where
the specular reflection rule is used (i.e., mirror like).
This implies that the incidence angle is equal to the
reflection angle upon collision: for a general reference, see, e.g.,
\cite{ref1}. The first studies about billiards began with Birkhoff 
in 1929 \cite{Birkhoff}, who proposed the billiard ball
motion in a manifold with an edge. After the pioneering results of Sinai
\cite{ref3}, Bunimovich \cite{ref6,ref7} and Gallavotti \cite{galavoti},
who gave mathematical support to the field, several applications of
billiards have been found in different areas of research including
optics \cite{ref8,ref9,ref10,edl}, quantum dots \cite{ref11}, microwaves
\cite{ref12}, astronomy \cite{ref13}, laser dynamics \cite{ref14} and
many others.

In the case when the billiard boundary is time-dependent,
$\partial Q=\partial Q(t)$, depending on the shape of the border, the
particle can accumulate energy under the effect of successive
collisions, leading to the phenomenon known as Fermi acceleration (FA).
Introduced in 1949 by Enrico Fermi \cite{ref15} as an attempt to
explain the high energy of the cosmic rays, FA basically consists in the
unlimited energy growth of a point-like particle suffering collisions
with an infinitely heavy and time-dependent boundary. One of the
important questions which arises from the study of 2-D time-dependent
billiards is: {\it what is the condition which leads the particle to
experience unlimited energy growth?} As previously discussed in Ref.
\cite{ref21}, the Loskutov-Ryabov-Akinshin (LRA) conjecture claims that
the chaotic dynamics for a particle in a billiard with static boundary
is a sufficient condition to produce FA if a time-perturbation to the
boundary is introduced. Later, F. Lenz et al in Ref. \cite{ref22} studied the case of a specific
time-perturbation to the elliptical billiard, which is integrable for
the static boundary. In this elliptic particular case, FA is produced by orbits
that ``cross" in the phase space the region of the separatrix, which
marks the separation from motion of two kinds: (i) libration and (ii)
rotation; therefore characterizing a change in the dynamics from
librator to rotator \cite{ref1,ref22}, or vice-versa. This ``crossing"
behavior is indeed assumed to be the FA production mechanism. Latter on
\cite{ref23} was shown that when the crossings were stopped, the FA is
suppressed and the energy of the particle is constant for long time
dynamics. As illustrated in \cite{ref22} and confirmed in \cite{ref23}
for a different time-perturbation, the existence of a separatrix curve
in the phase space, which is observed in the static case, turns into a
stochastic layer when a time perturbation is introduced on the boundary,
and produces the needed condition for the particle to accumulate energy 
along its orbit leading the dynamics to exhibit FA.

While the condition to produce FA in billiards is well understood, a
question which naturally arises is: ``What should one do to suppress FA
in time-dependent billiards if the energy of the particle is
unlimited?" The study of suppression of FA is quite recent for
two-dimensional billiards and one can consider different kinds of
dissipative forces. To illustrate a few of them, the introduction of
inelastic collisions has been discussed for the oval-like billiard \cite{ref29},
Lorentz gas \cite{ref30} and elliptical billiard \cite{ref23}. The
introduction of a drag force can also be considered a way to suppress
FA, as demonstrated in the oval-like billiard \cite{NEW} and elliptical
billiard \cite{NEW1}.

In this paper, we consider the dynamics of a time-dependent stadium-like
billiard, aiming to understand and describe the behavior of the average
velocity for an ensemble of initial conditions when dissipation is
introduced. First, we construct the equations describing
the dynamics of the model, including those governing the reflection
rule. Then we investigate the dynamics for regimes of high and low energy.
The initial high energy regime, as will be shown,
decays in time according to a power law. Critical exponents are derived
and scaling arguments are used to describe a scaling invariance for the
average velocity in the regime of high energy. Generally the dynamics of
the system depends on the control parameters including those controlling
the non-linearity of the system. As they are changed, average quantities
of some observables exhibit typical behavior observed in phase
transitions \cite{ref36}. Near the phase transition, critical exponents
can be defined and a scaling investigation can be carried out. Our results
for the dissipative stadium-like billiard show that the phenomenon of
FA is suppressed even for small dissipative coefficients, then changing
the regime of unlimited energy growth to limited growth. Since the
conservative version of this billiard presents FA
\cite{losk1,losk2,losk3}, this result give support to the principle
that FA seem not to be a structurally stable phenomenon \cite{edl1}.

This paper is organized as follows: In Sec. \ref{sec2} we construct the
equations that describe the dynamics of a dissipative time-dependent
stadium-like billiard. Section  \ref{sec3} is devoted to discuss our
results and is therefore divided in two parts: (i) The first one
includes the investigation of the chaotic transient considering high and
low energy regimes, including the scaling investigation and critical
exponents. (ii) In the second part, the final convergence for the
velocity is studied as function of the dissipation and initial energy
regime. Our final remarks and conclusions are drawn in Sec. \ref{sec5}.

\section{The Model and the Mapping}
\label{sec2}

In this section we construct the equations that describe the dynamics
of the system. The model describes the dynamics of a
point-like particle suffering inelastic collisions with a time-dependent
stadium billiard. Inelastic collisions are introduced by two distinct
damping coefficients $\gamma\in[0,1]$ and $\beta\in[0,1]$, where
$\gamma$ corresponds to the restitution coefficient with respect to the
normal component of the boundary at the instant of the collision
while $\beta$ is the restitution coefficient with respect to the tangential
component. For $\gamma=\beta=1$, as expected, results of the
non-dissipative case are all obtained. To construct the dynamics of the
model, we have to consider two distinct situations: (i) successive
collisions and;  (ii) indirect collisions. For case (i), the particle
suffers successive collisions with the same focusing component. On the
other hand, in (ii), after suffering a collision with a focusing
boundary, the next collision of the particle is with the opposite one
where the particle can, in principle, collide many times with the
parallel borders. We have considered that the time dependence in the
boundary is $R(t)=R_0+r\sin(wt)$, where $R_0\gg r$. The velocity of the
boundary is obtained by
\begin{equation}
\dot{R}(t)=B(t)=B_0\cos(wt)~,
\label{eq1}
\end{equation}
with $B_0=rw$ and $r$ is the amplitude of oscillation of the moving
boundary while $w$ is the frequency of oscillation. In our approach, the
dynamics of the model is described in its simplified version in the
sense that the boundary is assumed to be fixed, but, at the moment of
the collisions, it exchanges energy with the particle as if it was
moving. We consider fixed $w=1$.

\begin{figure}[ht]
\centering{\includegraphics[width=1.0\linewidth]{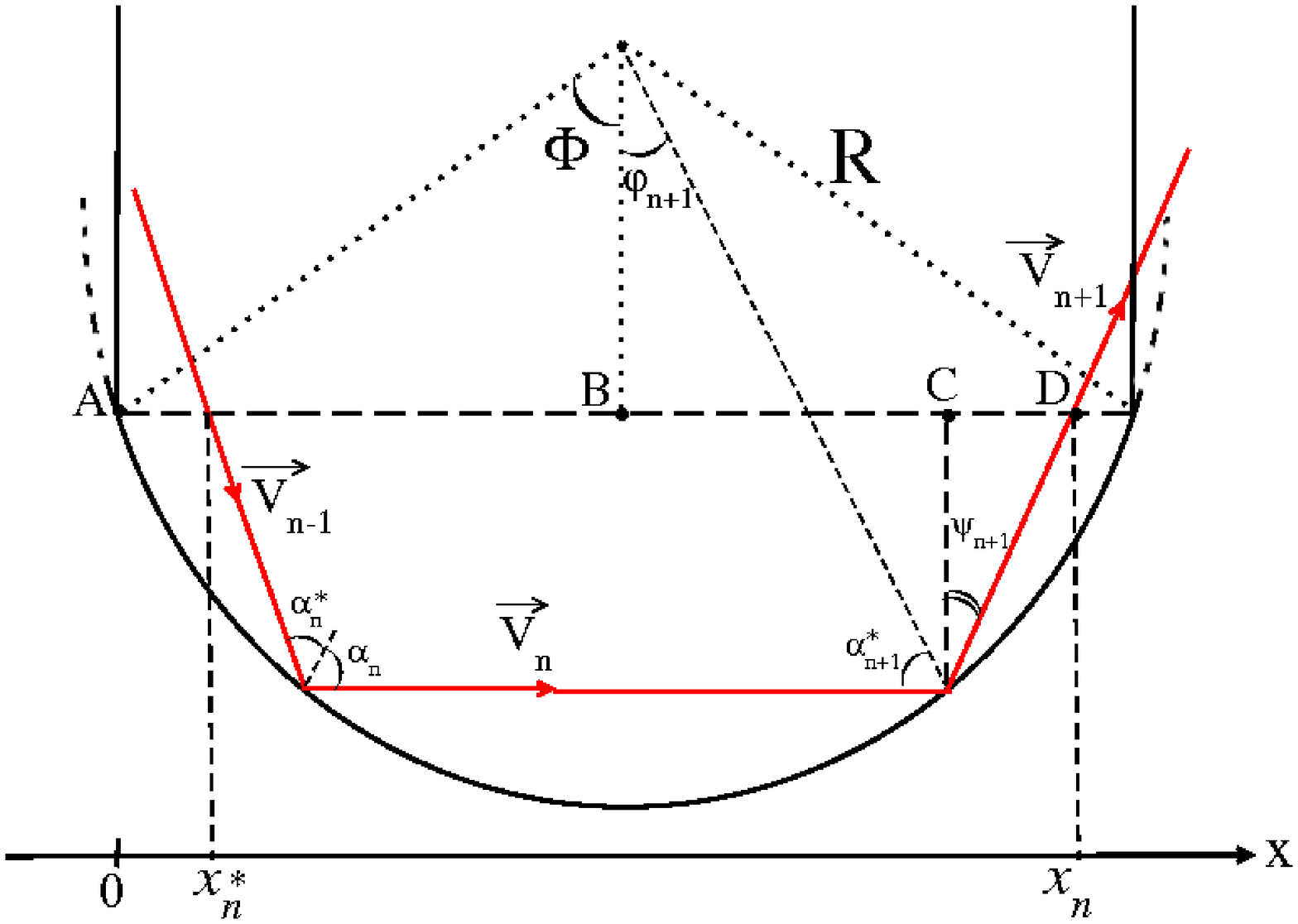}}
\caption{(color online) {Dynamical variables and a typical trajectory
with successive collisions.}}
\label{fig1}
\end{figure}

\begin{figure}[hb]
\centering{\includegraphics[width=1.0\linewidth]{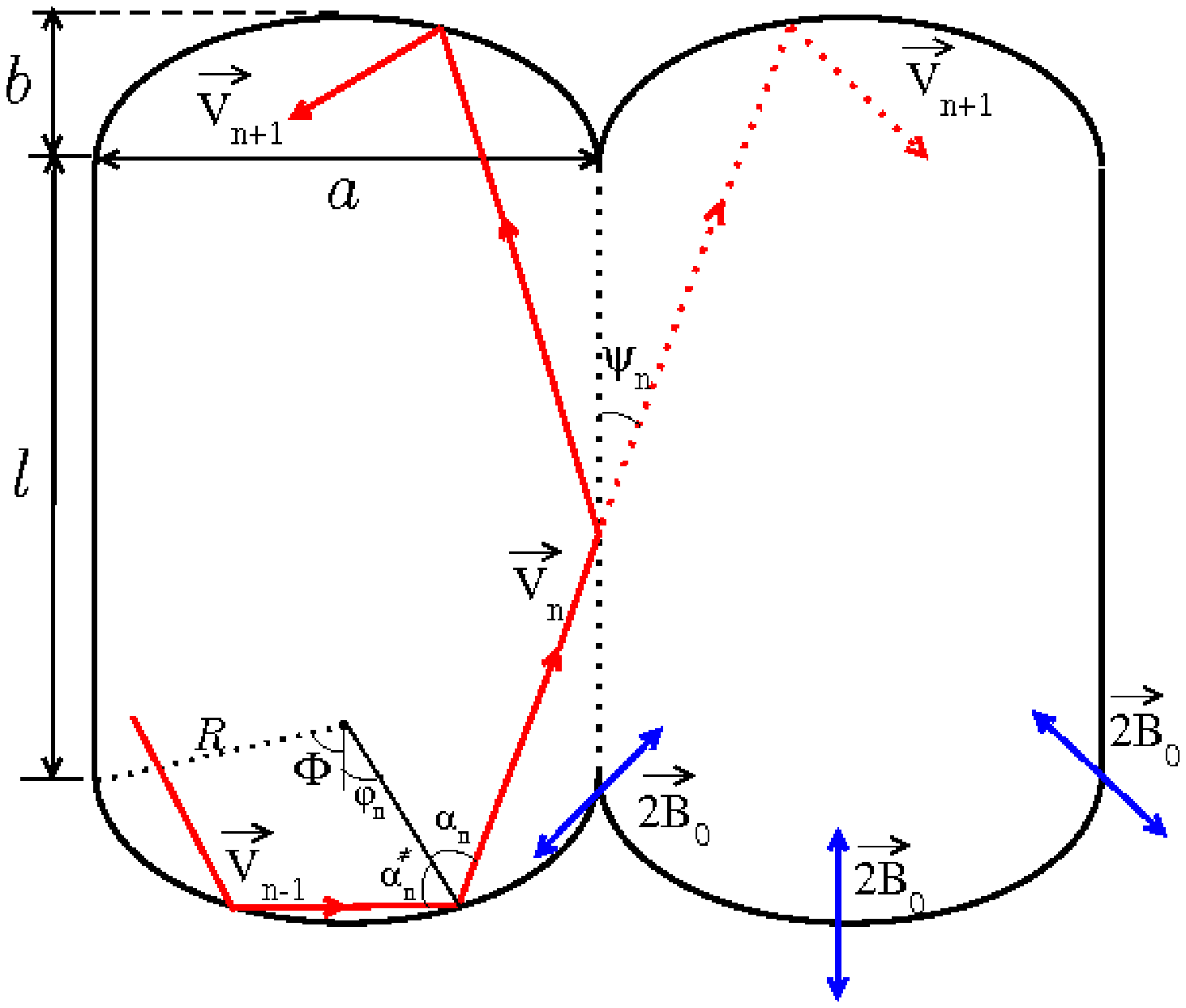}}
\caption{(color online) Illustration of the: (i) unfolding method; (ii)
control parameters and (iii) a typical trajectory with indirect
collisions with the focusing component.}
\label{fig2}
\end{figure}

The dynamics is evolved considering the variables
$(\alpha_n,\varphi_n,t_n,V_n)$, where $\alpha$ is the angle between the
trajectory of the particle and the normal line at the collision point,
$\varphi$ is the angle between the normal line at the collision
point and the vertical line in the symmetry axis. We assume that $V$
is the velocity of the particle and $t$ is the time at the instant of
the impact while the index $n$ denotes the $n^{th}$ collision of the
particle with the boundary. As an initial condition, we assume that at
the initial time $t=t_n$, the particle is at the focusing boundary and
the velocity vector directs towards to the billiard table. In the
notation, all variables with (*) are measured immediately before the
collision. Figure \ref{fig1}
shows an illustration of a trajectory with successive collisions, where
$R=(a^2+4b^2)/(8b)$ and $\Phi=\arcsin(a/2R)$. The control parameters
$a$, $b$ and $l$ are drawn in Fig.\ref{fig2}.
If $a=2b$ the original stadium billiard is recovered.

Let us consider case (i) first. For the occurrence of a successive
collision, it is necessary that $\mid\varphi_{n+1}\mid \leq \Phi$.
Therefore, according to Fig. \ref{fig1} and the specular reflection, we
have that
\begin{eqnarray}
\alpha^*_{n+1} &=& \alpha_n~, \nonumber \\
\varphi_{n+1} &=& \varphi_n +\pi -2\alpha_n~~ (mod~ 2\pi)~, \nonumber \\
t_{n+1} &=& t_n + {2R\cos(\alpha_n)\over V_n}~. \nonumber
\label{eq2}
\end{eqnarray}

Considering the case of indirect collisions, case (ii), it is necessary
that $\mid\varphi_{n+1}\mid>\Phi$. To obtain the equations describing
the dynamics, it is useful to consider the unfolding method
\cite{ref1}. Then two auxiliary variables, $\psi$, which is the angle
between the trajectory and the vertical line at the collision point and,
$x_n$, which is the projection of the particle position under the
horizontal axis are used. From geometrical considerations of Fig.
\ref{fig1} we obtain that $\psi_n=\alpha_n-\varphi_n$. One sees also
that $x_n$ is the summation of the line segments
$\overline{AB}+\overline{BC}+\overline{CD}$. Taking into account the
expression of $\psi_n$, and after some algebra, we obtain
$x_n={R\over\cos(\psi_n)}[\sin(\alpha_n)+\sin(\Phi-\psi_n)]$. The
recurrence relation between $x_n$ and $x_{n+1}$ is given by the
unfolding method, described in Fig. \ref{fig2}, as
$x_{n+1}=x_n+l\tan(\psi_n)$.

To obtain the angular equations, we invert the particle motion, i.e.,
consider the reverse direction of the billiard particle, then the
expression that furnishes $x_n$ is also inverted and the angle
$\alpha_n$ becomes $\alpha^{*}_n$. Resolving it with respect to
$\alpha^{*}_n$, taking into account that this angle is changed in the
opposite direction, then the angles $\alpha_n$ and $\varphi_n$ must have
their sign reversed. Moreover, the incident angle $\alpha^{*}_n$ assumes
$\alpha^{*}_{n+1}$ when the motion of the particle is re-inverted. The
expressions of $\varphi_{n+1}$ and the time $t_{n+1}$ are obtained by
simple geometrical considerations of Fig. \ref{fig2}. Thus, we obtain
the mapping for the case of indirect collisions as
\begin{eqnarray}
\alpha^{*}_{n+1}&=&\arcsin[\sin(\psi_n+\Phi)-x_{n+1}\cos(\psi_n)/R]~, 
\nonumber  \\
\varphi_{n+1}&=&\psi_n-\alpha^*_{n+1}~, \nonumber \\
t_{n+1}&=&t_n+{R[\cos(\varphi_n)+\cos(\varphi_{n+1})-2\cos(\Phi)]+l\over
V_n\cos(\psi_n)}~. \nonumber
\label{eq3}
\end{eqnarray}

For both cases (i) and (ii), the recurrence relations for $V_n$ and
$\alpha_n$ are the same. Let us discuss how to obtain them. Expressing
the two components of the velocity vector before the collisions we
have that
\begin{eqnarray}
\vec{V}_n\cdot\vec{T}_{n+1}&=&-V_n\cos(\alpha^{*}_n)~, \nonumber \\
\vec{V}_n\cdot\vec{N}_{n+1}&=&V_n\sin(\alpha^{*}_n)~. \nonumber
\label{eq4}
\end{eqnarray}
Given that the moving boundary is not an inertial referential frame, we
assume that at the instant of the collision, the wall is instantaneously
at rest, then the reflection laws are given by
\begin{eqnarray}
\vec{V}^{\prime}_{n+1}\cdot\vec{T}_{n+1}&=&\beta\vec{V}^{\prime}
_n\cdot\vec { T }_{n+1}~, \nonumber \\
\vec{V}^{\prime}_{n+1}\cdot\vec{N}_{n+1}&=&-\gamma\vec{V}^{\prime}
_n\cdot\vec { N }_{n+1}~, \nonumber
\label{eq6}
\end{eqnarray}
where $\beta$ and $\gamma$ are respectively the restitution
coefficients with respect to the tangent and the normal components of
the motion. We stress that $\vec{V}^{\prime}_{n+1}\cdot\vec{T}_{n+1}$
and $\vec{V}^{\prime}_{n+1}\cdot\vec{N}_{n+1}$ are the components of the
velocity of the particle measured in the referential frame of the
moving wall with respect to the tangent and the normal components
respectively.

In the inertial referential frame, we have that the equations for the
components of the velocity are given by
\begin{eqnarray}
\vec{V}_{n+1}\cdot\vec{T}_{n+1}=\beta\vec{V}_n\cdot\vec{T}_{n+1}
+(1-\beta)\vec { B }(t_{n+1})\cdot\vec{T}_{n+1}~, \nonumber  \\
\vec{V}_{n+1}\cdot\vec{N}_{n+1}=-\gamma\vec{V}_n\cdot\vec{N}_{n+1}
+(1+\gamma)\vec {B}(t_{n+1})\cdot\vec{N}_{n+1}~, \nonumber
\label{eq7}
\end{eqnarray}
where $B(t_{n+1})$ is obtained from Eq.(\ref{eq1}) evaluated at the
time $t_{n+1}$. Finally, the expression of $V_{n+1}$ is given by
\begin{equation}
\mid \vec{V}_{n+1}\mid =
\sqrt{(\vec{V}_{n+1}\cdot\vec{T}_{n+1})^2+(\vec{V}_{n+1}\cdot\vec{N}_{
n+1})^2}~.
\label{eq8}
\end{equation}

The last equation refers to the reflection angle, which according to
the reflection law is given by
\begin{equation}
\alpha_{n+1}=\arcsin\left({\mid\vec{V}_n\mid \over \mid \vec{V}_{n+1}
\mid} \sin(\alpha^{*}_{n+1})\right)~.
\label{eq9}
\end{equation}

\section{Discussions of the Dynamics and numerical results}
\label{sec3}

To investigate the dynamics of the model, we set as fixed the parameters
$a=0.5$, $b=0.01$, $l=1$ and the amplitude of oscillation of the moving
wall as $B_0=0.01$. The reason for keeping the parameters fixed is
because some observables are scaling invariant with respect to the
control parameters, as discussed for the static case in Ref. \cite{liv}.
Therefore for this paper we chose to investigate the dynamics by the
variation of the parameter $\gamma$ as well as the initial velocity
$V_0$. The parameter $\beta$ is considered fixed as $\beta=1$. Moreover
we stress that collisions of the particle with the straight segments of
the border are considered elastic.

\begin{figure}[t]
\centering{\includegraphics[width=1.0\linewidth]{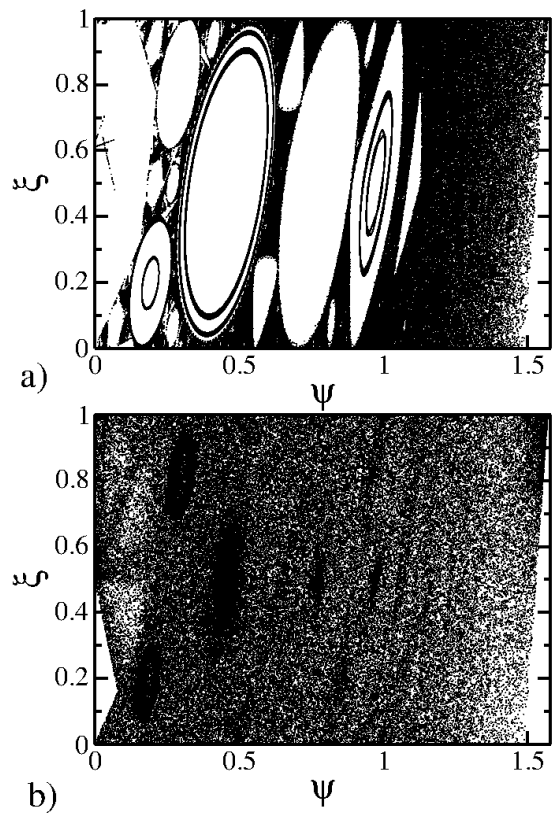}}
\caption{Phase portrait of $(\psi,\xi)$ for: (a) $V_0=10$,
$\gamma=1.0$ and (b) $V_0=10$, $\gamma=0.999$.}
\label{fig_espaco_psi_qsi}
\end{figure}

Figure \ref{fig_espaco_psi_qsi} illustrates the dynamics of the particle
on the variables $(\psi,\xi)$, where $\xi =x/a$, for the conservative case and the dissipative dynamics respectively.
It is known from the literature
that, when dissipation is introduced in the dynamics, the invariant
curves surrounding the fixed points might be destroyed and the elliptic
fixed points turn into sinks \cite{litch}. Specific discussions of fixed
points in the non dissipative case can be seen for the stadium-like
billiard in Refs. \cite{losk1,losk2,losk3,liv}. The procedure used to
construct the phase portraits was to consider 25 different initial
conditions and evolve each initial condition until $10^7$ collisions.
For the conservative case, FA is indeed oberserved for the stadium billiard for an initial velocity higher 
than the critical resonant one \cite{losk1}, and the phase space for this unlimited energy growth is shown in Fig.\ref{fig_espaco_psi_qsi}(a).
On the other hand, the darker regions of Fig. \ref{fig_espaco_psi_qsi}(b) mark the
convergence to the attractors while the spread points along the plot
identify the {\it seemingly} chaotic transient. The control parameters
and initial conditions used in the construction of the figure were: (a)
$V_0=10$, $\gamma=1.0$ and (b) $V_0=10$, $\gamma=0.999$. Comparing Fig.\ref{fig_espaco_psi_qsi}(a) and (b), one can see the ``birth'' of a period-2 atractor, 
and two period one atracting ponits.

In order to have a better understanding of the dynamics, particularly
the convergence of the initial conditions to the attractors, we
concentrate to study the dynamics and hence some properties of the
average velocity of the particle. We consider dependence of the average
velocity as function of both: (a) number of collisions with the boundary
$n$; (b) initial velocity $V_0$ and; (c) restitution coefficient
$\gamma$. The average velocity is therefore defined as
\begin{equation}
\overline{V}= {1\over M} \sum_{j=1}^M {V_{i,j}(n,V_0,\gamma)}~,
\label{eq10}
\end{equation}
where $M$ is an ensemble of $5000$ different initial conditions
$(\alpha,\varphi)$, from which $\xi$ and $\psi$ are described, and $V_i$ is expressed by
\begin{equation}
V_i(n,V_0,\gamma)= {1\over n} \sum_{k=1}^n {V_k}~,
\label{eq11}
\end{equation}
where $n$ is the number of collisions with the moving wall.

The numerical investigations were carried out in two different ways.
To understand the behavior of the average velocity and hence the
energy of the particle at the range of large initial velocity we
consider two regimes of time: (i) short time, mainly marked by the
dynamics evolving through a transient and; (ii) long time where the
dynamics has already reached the attractor. For high initial velocity,
the particle experiences a decay in the average velocity marked by a
transient in the dynamics. Moreover the decay of energy is described by
a homogeneous function with critical exponents. (ii) For long time, a
statistics is made to the regime of convergence ir order to understand 
the role of the dissipation and of the initial energy regime. The simulations were
evolved up to $n=5\times 10^8$ collisions and considering an ensemble of
$5000$ different initial configurations uniformly chosen as $\varphi
\in [0,\Phi]$ and $\alpha \in [0,\pi/2]$. Due to the axial symmetry of
the stadium-like billiard, a negative range of the initial
conditions is not needed.

\subsection{Transient for short time}
\label{subsec1}

\begin{figure}[ht]
\centering{\includegraphics[width=1.0\linewidth]{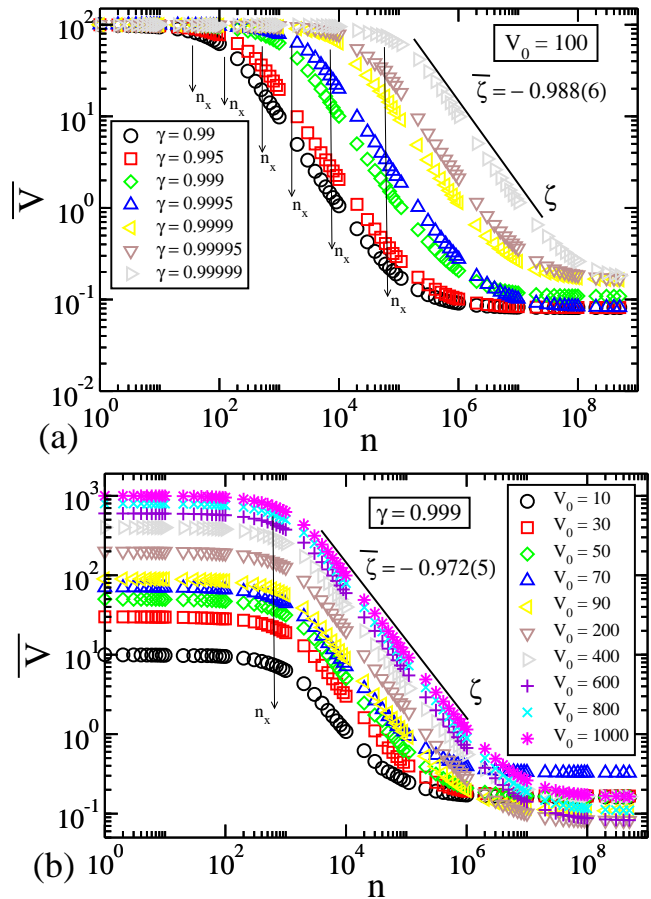}}
\caption{(color online) Plot of $\overline{V}~vs~n$ for the large
initial velocity. The parameters and initial conditions used were: (a)
$V_0=100$ and different restitution coefficients $\gamma$ and; (b) 
different initial velocities and a fixed $\gamma=0.999$.}
\label{fig3}
\end{figure}

In this section we concentrate to investigate the initial transient.
Therefore we consider two ranges of initial velocity: (i) large and
(ii) low. Let us start with the high energy. Figures \ref{fig3}(a,b)
illustrate the behavior of $\overline{V}$ as function of the number of
collisions. In Fig. \ref{fig3}(a) we assume as fixed the initial
velocity as $V_0=100$, and varied the restitution coefficient $\gamma$.
In Figure \ref{fig3}(b) we considered fixed $\gamma=0.999$ and varied
the initial velocity $V_0$. The range of $V_0$ was chosen in such a way
to configure a very large initial velocity as compared to the maximum
boundary velocity $V_0>>r\omega$. The curves shown in Figs.
\ref{fig3}(a) and \ref{fig3}(b) indeed exhibit similar behavior for
short time. They begin in a constant regime for each initial velocity
and suddenly, depending on the value of the damping coefficient
$\gamma$, they experience a crossover $n_x$ marking a change from a
constant regime and bend towards a regime of decay according a power
law. A careful fitting in the curves gives that the decay exponent is
$\zeta\approx -1$. After the decay observed for large $n$, the curves of
$\overline{V}$ saturate in different plateaus of low energy, which may
depend of both $V_0$ and $\gamma$. The plateaus characterize indeed
different attractors to where the dynamics has converged to.
For large enough time the convergence regions are around the
range $V\in (0.07, 0.6)$ The investigation of these plateaus will be
made in the next subsection.

Given that the initial behavior of $\bar{V}$ for the range of large
initial $V_0$ is similar even for different values of $V_0$ and
$\gamma$, we can suppose that:
\begin{itemize}
\item{
{(i)} $\overline{V} \propto V_0^{\alpha}$, for $n\ll n_x$, where
$\alpha$ is a critical exponent and $n_x$ is the characteristic crossover collision;}
\item{
{(ii)} $\overline{V} \propto \left({n \over V_0}\right)^{\zeta}$, for
$n\gg n_x$, where $\zeta\approx -1$ is the power law decaying
exponent;}
\item{
{(iii)} $\left({n_x\over V_0}\right)\propto V_0^{z_1}(1-\gamma)^{z_2}$,
where $z_1$ and $z_2$ are dynamical exponents.}
\end{itemize}

\begin{figure}[ht]
\centering{\includegraphics[width=1.0\linewidth]{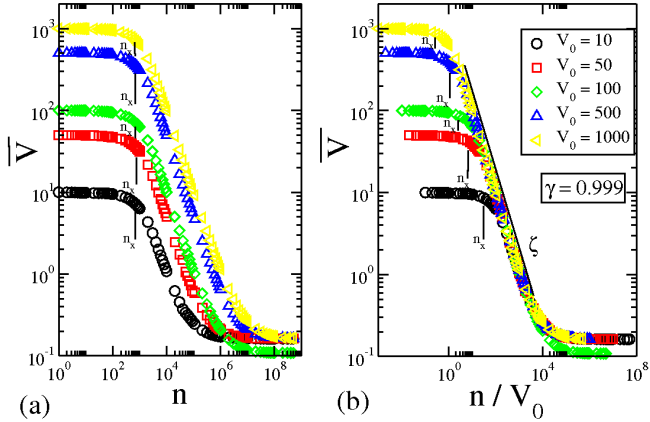}}
\caption{(color online) Plot of $\overline{V}$ as a function of: (a) $n$
and (b) $n/V_0$. The parameter $\gamma$ was fixed as $\gamma=0.999$ and
5 different initial conditions were used, as labeled in the figure.}
\label{fig7}
\end{figure}

On the scaling hypothesis (iii) we considered $(1-\gamma)$ instead
of $\gamma$, because we want to consider the transition
$(1-\gamma)\rightarrow 0^+$. We see also that different initial
velocities produce different curves of $\overline{V}$ but with the same
negative slope. Therefore the transformation $n\rightarrow n/V_0$ makes
all the curves coincide in the decay regime. This transformation together with
some curves of $\overline{V}$ are shown in Fig.
\ref{fig7}.
After this transformation, all curves of $\overline{V}$ start in a
constant regime and then decay together as a power law with exponent
$\zeta\approx -1$.

\begin{figure}[ht]
\centering{\includegraphics[width=1.0\linewidth]{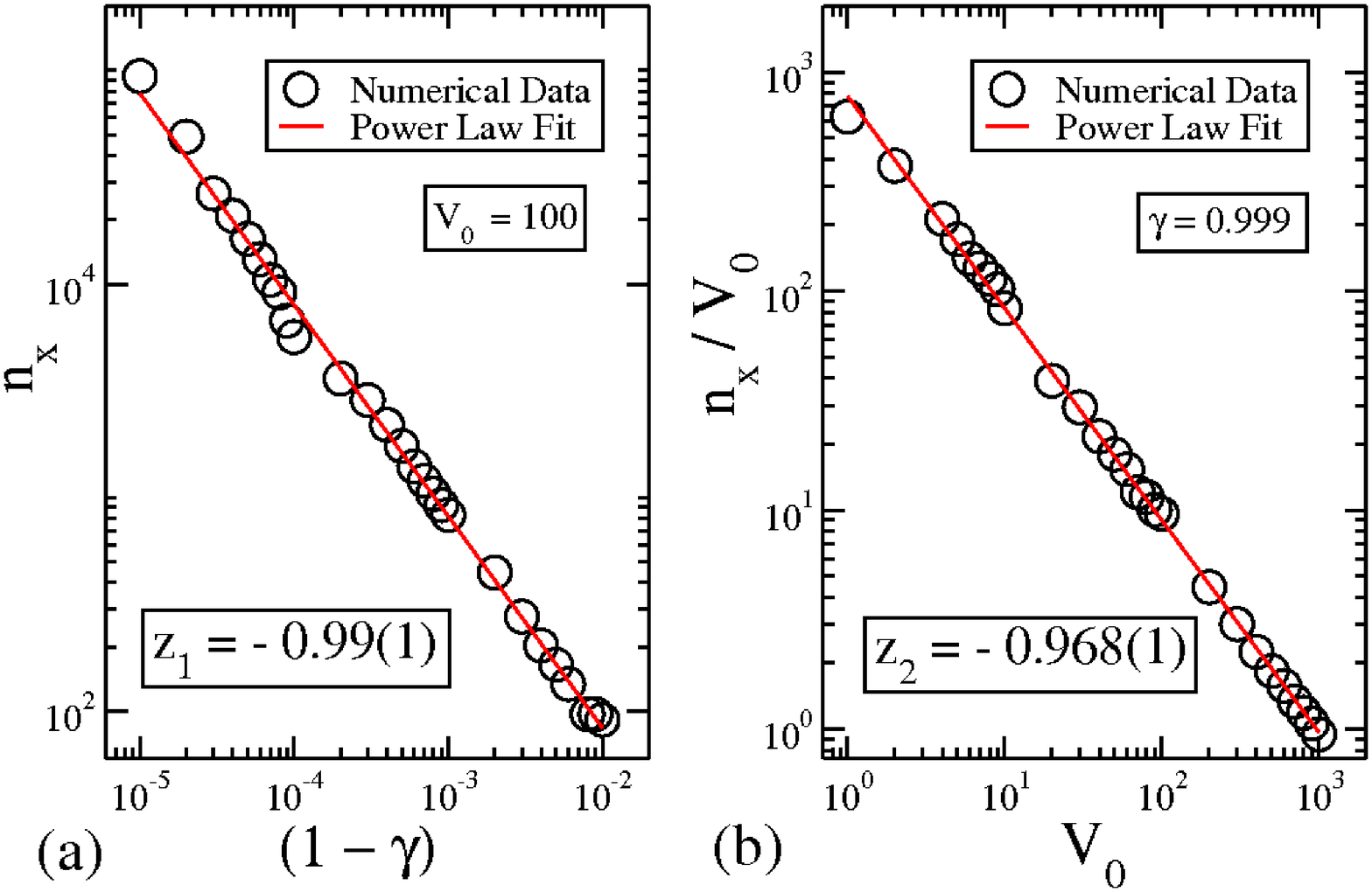}}
\caption{(color online) Plot of: (a) $n_x/V_0~vs~V_0$ for $\gamma=0.999$
and; (b) $n_x\times (1-\gamma)$ for $V_0=100$. After fitting the
data we obtain $z_1=-0.99(1)$ and $z_2=-0.968(1)$.}
\label{fig8}
\end{figure}

Given that the initial behavior of $\overline{V}$ is constant, we
conclude that $\alpha=1$. The critical exponents $z_1$ and $z_2$ are
obtained respectively by power law fits on the plots
$(n_x/V_0)\times V_0$ and $(n_x/V_0)\times(1-\gamma)$. Figure \ref{fig8}
gives us that $z_1=-0.99(1)$ and $z_2=-0.968(1)$.

The three scaling hypotheses allow us to describe the behavior of
$\overline{V}$ for short $n$ (before the convergence to the constant
plateau) formally as a scaling function of the type
\begin{equation}
\overline{V}(n/V_0,V_0(1-\gamma))=\lambda\overline{V}(\lambda^{a_1}n/V_0
, \lambda^ { b_
1} V_0(1-\gamma))~ ,
\label{eq12}
\end{equation}
where $\lambda$ is a scaling factor and $a_1$ and $b_1$ are scaling
exponents. Assuming that $\lambda^{a_1} n/V_0$ constant, we have
\begin{equation}
\lambda=\left({n\over V_0}\right)^{-{1\over a_1}}~.
\label{eq13}
\end{equation}

Substituting Eq.(\ref{eq13}) in (\ref{eq12}), we obtain
\begin{equation}
\overline{V}(n/V_0,V_0(1-\gamma))=V_0^{-{1\over a_1}}
\overline{V_1}(1,\lambda^{-{b_1\over a_1}}V_0(1-\gamma))~.
\label{eq14}
\end{equation}
If we compare Eq.(\ref{eq14}) with the hypothesis {(ii)} and assuming
that $\overline{V_1}$ is constant for $n\gg n_x$, we obtain
\begin{equation}
\zeta=-{1\over a_1}~,
\label{eq15}
\end{equation}
and given that the critical exponent $\zeta\approx-1$, obtained by
fitting a power law to the decay regime, we have that $a_1=1$.\

Choosing now $\lambda^{b_1} V_0(1-\gamma)$ constant, we have
\begin{equation}
\lambda=\left(V_0(1-\gamma)\right)^{-{1\over{b_1}}}~.
\label{eq16}
\end{equation}
Replacing Eq.(\ref{eq16}) in (\ref{eq12}), we obtain
\begin{equation}
\overline{V}(n/V_0,V_0(1-\gamma))=V_0(1-\gamma)^{-{1\over{b_1}}}
\overline{V_2}(\lambda^{-{{a_1}\over{b_1}}}n/V_0,1)~.
\label{eq17}
\end{equation}

A comparison of Eq.(\ref{eq17}) with hypothesis {(i)}, and assuming
$\overline{V_2}$ constant for $n\ll n_x$, leads to
\begin{equation}
\alpha=-{1\over b_1}~,
\label{eq18}
\end{equation}
and given the constancy of the initial regime, we conclude that
$\alpha=1$, yielding $b_1=-1$.\

Comparing now Eqs.(\ref{eq13}) with (\ref{eq16}) and after 
straightforward algebra, we obtain that
\begin{equation}
\left({n\over V_0}\right)=\left(V_0(1-\gamma)\right)^{{a_1\over b_1}}~.
\label{eq20}
\end{equation}
When Eq. (\ref{eq20}) is compared with the scaling hypothesis
{(iii)} we conclude that
\begin{equation}
{a_1\over b_1}={\alpha\over \zeta}=z_1=z_2=-1~.
\label{eq21}
\end{equation}

\begin{figure}[ht]
\centering{\includegraphics[width=1.0\linewidth]{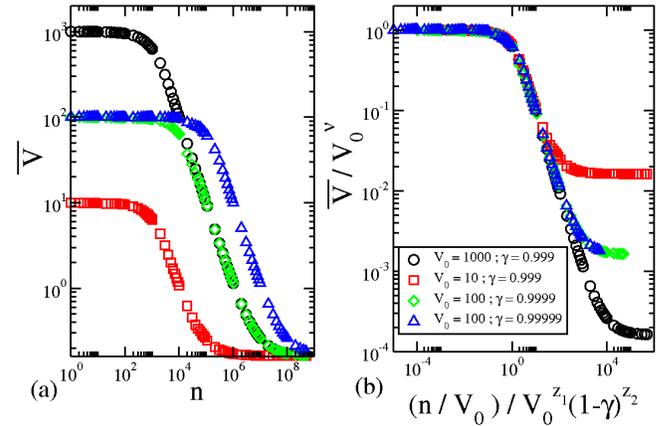}}
\caption{(color online) Plot of: (a) $\overline{V}~vs~n$, and (b)
overlap of the initial transient of all curves of (a) onto a single
plot, after a suitable rescale of the axis.}
\label{fig9}
\end{figure}

This procedure and the critical exponents let us rescale properly both
axis of the $\overline{V}~vs~n$ plot and obtain a single and universal
curve for the short time transient dynamics, as shown in Fig.
\ref{fig9}.
Basically, this result confirms that, independent of the initial
velocity and the control parameter $\gamma\cong 1$, the behavior of the
transient for the high energy regime of $\overline{V}$ curves is scaling
invariant with respect to $V_0$ and $\gamma$.

\begin{figure}[ht]
\centering{\includegraphics[width=1.0\linewidth]{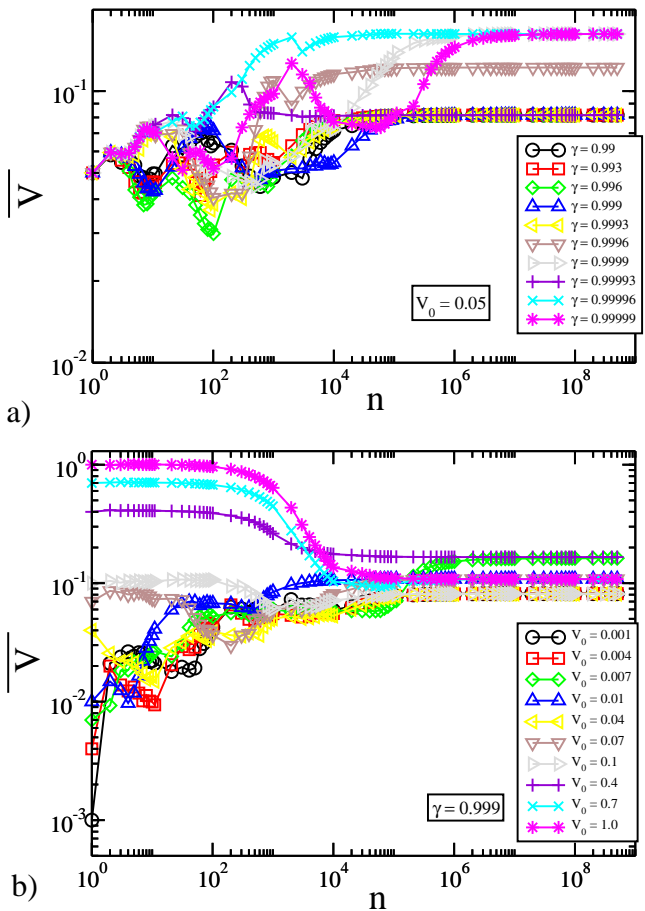}}
\caption{(color online)  Plot of $\overline{V}~vs~n$ for the low energy
regime: In (a) $V_0=0.05$ was fixed and the dissipation parameter
$\gamma$ was ranged while; (b) $\gamma=0.999$ was fixed and the initial
velocity for the a initial energy regime was ranged.}
\label{fig_baixa}
\end{figure}

Let us now consider the case where the initial velocity is low,
therefore the behavior of the transient is different. The curves of
$\bar{V}$ exhibit a regime of growth until they reach the convergence
regions, as shows Figs. \ref{fig_baixa}(a,b). The combination of control
parameters $V_0$ was set in order to keep the initial velocity in
the low energy regime. The dissipation and the initial velocity are
labeled in the figure.

The presence of the attractors indeed define the region to where the
curves of $\bar{V}$ converge to and therefore saturate. Although one
may think this is quite a paradoxical behavior in the sense that
dissipation leads to a regime of growth, if one looks deeper at the
dynamics, indeed the particle is only converging to an attractor wich
is located at higer energy compared to the initial velocity.
Considering that this attractor is not at infinite velocity, the FA is
suppressed.

\subsection{Sinks, attractors and Convergence of $\overline{V}$}

Once the behavior of the chaotic transient for both high and low energy
regimes is described, let us concentrate our efforts to investigate the
convergence regions, the role of the sinks and their dependence
according the dissipation and initial velocities.

\begin{figure}[ht]
\centering{\includegraphics[width=1.0\linewidth]{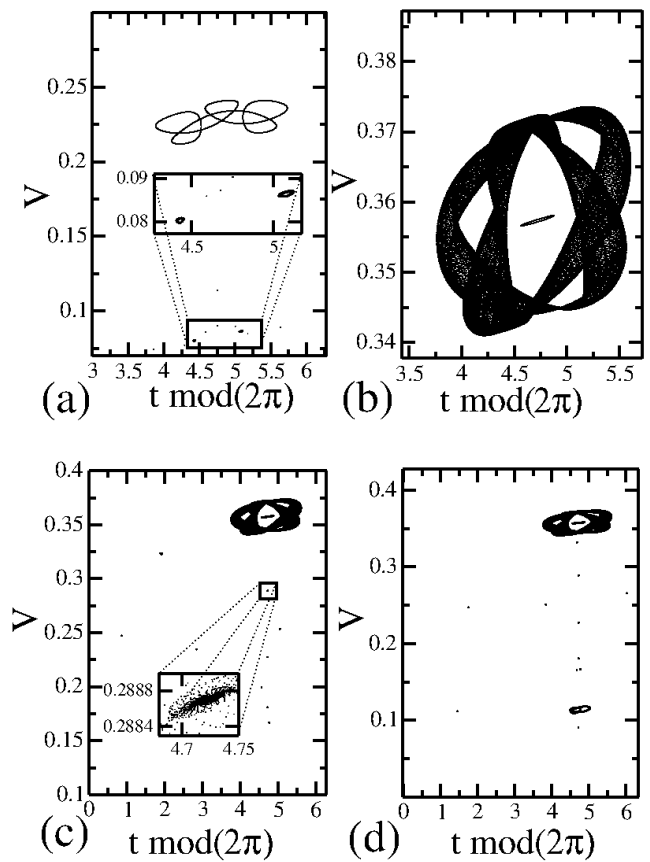}}
\caption{Plot of $V\times t~mod(2\pi)$ after the chaotic transient for:
(a) $V_0=1000$, $\gamma=0.99$; (b) $V_0=100$, $\gamma=0.9999$; (c)
$V_0=100$, $\gamma=0.999$ and (d) $V_0=10$, $\gamma=0.999$.}
\label{fig6}
\end{figure}

As shown in Fig. \ref{fig_espaco_psi_qsi}, the fixed points in the space
$(\psi,\xi)$ become sinks. However a visualization of the phase
portraits in such a variables does not give any conclusions regarding the
final velocity. Then it is natural to look at the plots of $V$ {\it vs}
$t~mod(2\pi)$ where $t$ is the time shown in Fig. \ref{fig6}.
We can see that the convergence regions for the curves of $\overline{V}$
illustrated in Figs. \ref{fig3} and \ref{fig_baixa} are around
$V\in (0.07,0.6)$. It is also possible to see from Fig. \ref{fig6} many
different sets of attracting fixed points and more complex attractors. A
zoom-in window shows better some of the sinks in Figs. \ref{fig6}(a,c).
In particular, one can enumerate in Fig. \ref{fig6}(d) at least $13$
different attractors. Each attractor of this set produces a different
plateau in the asymptotic curves of $\bar{V}$. This multitude of attracting
points for the convergence zone is the reason why a scaling treatment
for long time is indeed a real challenge. To construct the figures, we
set $100$ different initial conditions, each one of them evolved in time
until $5\times 10^7$ collisions.
    
\begin{figure}[ht]
\centering{\includegraphics[width=1.0\linewidth]{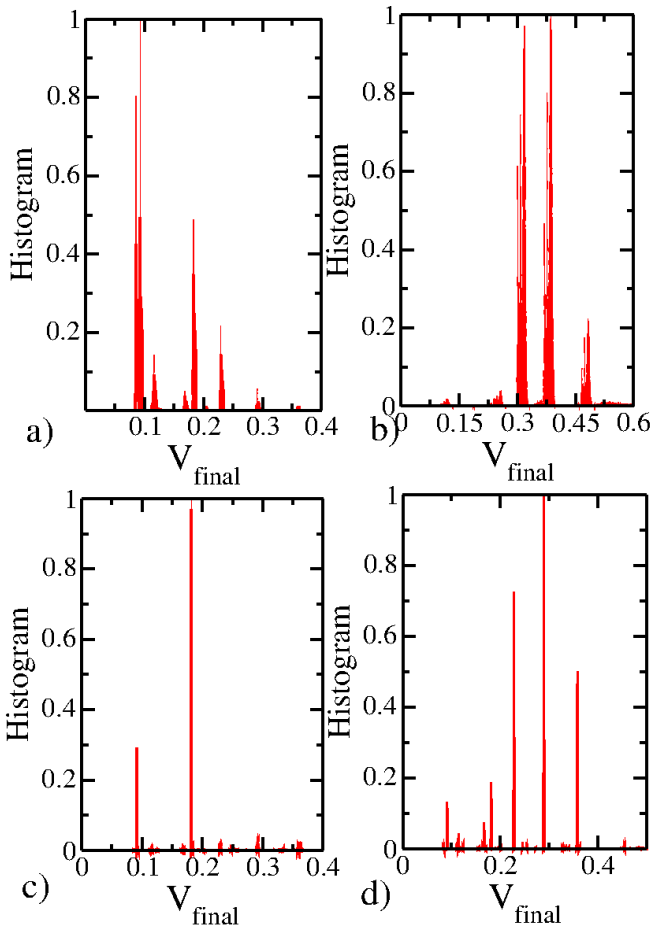}}
\caption{(color online) Histogram of frequency for the convergence
velocity region for: (a) $V_0=1000$, $\gamma=0.99$; (b) $V_0=100$,
$\gamma=0.9999$; (c) $V_0=100$, $\gamma=0.999$ and (d) $V_0=10$,
$\gamma=0.999$.}
\label{fig_histo}
\end{figure}

Each attractor has its own influence over the dynamics which is
dependent on the size of the basin of attraction. A way to see this is
constructing a histogram of frequency of initial conditions showing
convergence to a particular attractor. Figure \ref{fig_histo} shows the
corresponding histogram of frequency of visited attractors for the same
control parameter used in Fig. \ref{fig6} however with a large ensemble
of initial conditions, indeed $25\times 10^4$ of them where each one
of them was evolved until $5\times 10^8$ collisions. 

Figure \ref{fig_histo}(a) shows that the most visited attractors are
those located around $V\in(0.08,0.09)$. A comparison of this result with
Fig. \ref{fig6}(a) shows one main attractor indicating a possibly
period-2 attractor. Figure \ref{fig_histo}(b) shows that the
distribution of final velocity of the orbits, for the combination of
control parameters $\gamma=0.9999$ and $V_0=100$, are more concentrated
in a range $V\in (0.3,0.4)$. Again, if the results are compared with
Fig. \ref{fig6}(b), one sees that the visited region $V\in (0.3,0.4)$
corresponds to the seemingly cyclic attractor. Also, Figs.
\ref{fig_histo}(c,d) show many different visited regions corresponding
to the regions of final velocity as shown in Figs. \ref{fig6}(c,d).

\begin{figure}[ht]
\centering{\includegraphics[width=1.0\linewidth]{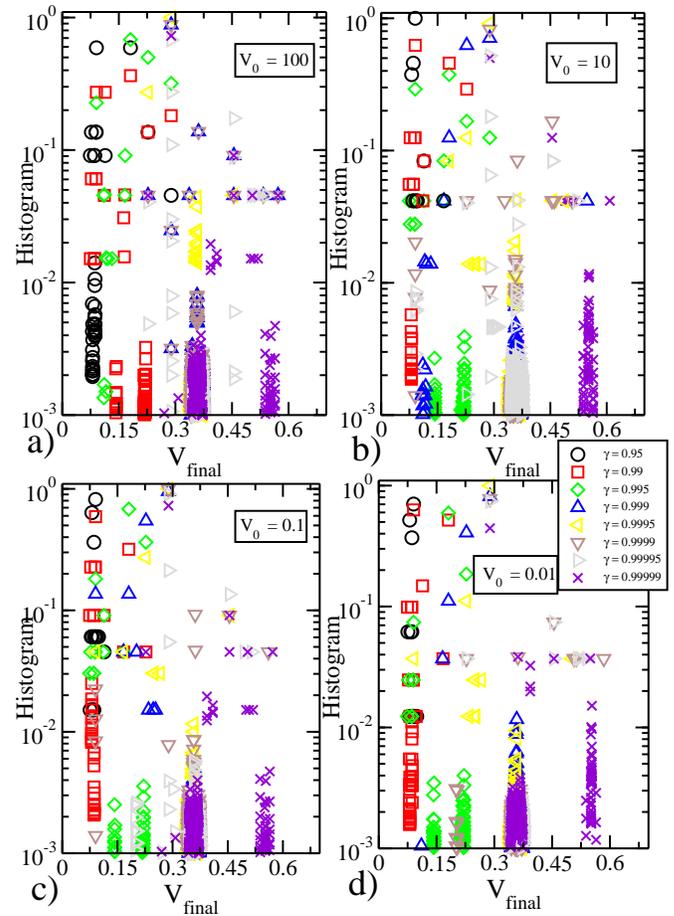}}
\caption{(color online) Histogram of frequency for different parameters
$\gamma$. The control parameters and initial condition used were: (a)
$V_0=100$, (b) $V_0=10$, (c) $V_0=0.1$ and (d) $V_0=0.01$.}
\label{fig_atratores}
\end{figure}

The influence of the attractors are dependent on the control parameters
used. Therefore specific attractors can be more influence than others
for a combination of control parameters. In order to understand the
dependence of the attractors with the dissipation parameter and the
initial velocity, we constructed a histogram of frequency taking into
account different control parameters and initial velocities, as shown
in Fig. \ref{fig_atratores}. After a careful look at Fig.
\ref{fig_atratores}, we conclude that the attractors for ``more''
dissipative systems, for example the ones for $\gamma\in[0.95,0.999]$,
prefers the regions of lower velocities even considering the high and
low initial velocity. On the other hand the ``less" dissipative
dynamics, for example for the range
$\gamma\in[0.9995,0.999999]$, prefers regions of higher velocities as
compared to the previous case. Each histogram shown in Fig.
\ref{fig_atratores} was constructed considering $25\times 10^4$
different initial conditions where each one of them was evolved up to
$5\times10^8$ collisions. Figures \ref{fig_atratores}(a,b) represent
the histograms for the initial high velocity while Figs.
\ref{fig_atratores}(c,d) show the histograms for the low initial
velocity. In particular, one can notice that in Fig.
\ref{fig_atratores}(d), the legend used for each parameter of
dissipation $\gamma$, applies to all plots of Fig. \ref{fig_atratores}.

\section{Conclusions}
\label{sec5}

In summary, we considered a time-dependent stadium-like billiard with
dissipation introduced via inelastic collisions. With the introduction
of the dissipation, we have shown that FA is suppressed. High initial
velocities, after a crossover time, decay as a power law of the number
of collisions with the border, while low initial energy regimes, lead
the particle to present a regime of ``growing" to the same convergence
region of the orbits with a high energy chaotic transient. This chaotic
transient for the high energy regimes, was characterized as function of
both $V_0$ and $\gamma$. Scaling arguments were used to overlap the
behavior of $\overline{V}$ for short $n$, showing that the dynamics
of short time is scaling invariant with respect to $V_0$ and $\gamma$,
if we considered high initial energy regime. The system is shown to
have many attractors, where some of them are sinks and others are more
complicate. When the damping coefficient is varied there is a
tendency for more dissipation bringing the dynamics to visit more
often the region $V_{final}\approx 0.30$. On the other hand, less
dissipative dynamics prefers the attractors around
$V_{final}\in(0.3,0.6)$. Finally, it is clear that the unlimited energy
growth is interrupted with the presence of inelastic collisions
therefore leading to one more example where Fermi acceleration seems not to
be a structurally stable phenomenon.

\section*{ACKNOWLEDGMENTS}
A.L.P.L thanks to FAPESP and CNPq for the financial support. I.L.C. also thanks FAPESP and CNPq. E.D.L
thanks to FAPESP, CNPq and Fundunesp, Brazilian agencies. This research
was supported by resources supplied by the Center for Scientific
Computing (NCC/GridUNESP) of the S\~ao Paulo State University (UNESP). The authors also thank Carl Dettmann for a careful reading on the manuscript.

\end{document}